# Tutorial: Neuromorphic spiking neural networks for temporal learning

*Doo Seok Jeong**

Division of Materials Science and Engineering, Hanyang University, 222 Wangsimni-ro, Seongdong-gu, Seoul 04763, Republic of Korea

E-mail: dooseokj@hanyang.ac.kr

**Abstract**

Spiking neural networks (SNN) as time-dependent hypotheses consisting of spiking nodes (neurons) and directed edges (synapses) are believed to offer unique solutions to reward prediction tasks and the related feedback that are classified as reinforcement learning. Generally, temporal difference (TD) learning renders it possible to optimize a model network to predict the delayed reward in an *ad hoc* manner. Neuromorphic SNNs—networks built using dedicated hardware—particularly leverage such TD learning for not only reward prediction but also temporal sequence prediction in a physical time domain. In this tutorial, such learning in a physical time domain is referred to as *temporal learning* to distinguish it from conventional TD learning-based methods that generally involve algorithmic (rather than physical) time. This tutorial addresses neuromorphic SNNs for temporal learning from the scratch. It first concerns general characteristics of SNNs including spiking neurons and information coding schemes and then moves on to temporal learning including its general concept, feasible algorithms, and their association with neurophysiological learning rules that have intensively been enriched for the last few decades.

## I. Introduction

Neurogrid[1], TrueNorth[2], DYNAPs[3], SpiNNaker[4], Loihi[5], and Braindrop[6] are prototypical neuromorphic processors that have recently been released. They are designed using very distinct technologies so that their working principles and capabilities all differ. Nevertheless, it is noticed that they all realize spiking neural networks (SNNs) on the chips, the realization techniques are significantly different though. In fact, it is not surprising that the aforementioned neuromorphic prototypes are commonly based on SNNs with regard to the three decades of history of neuromorphic engineering.[7,8] Neuromorphic engineering initially aimed to reverse-engineer the central nervous system (CNS) including the retina. Given that SNNs can model the CNS, SNNs are the network type of significant concern in neuromorphic engineering.

SNN is a dynamic hypothesis in which neuronal information is conveyed from a presynaptic neuron to a postsynaptic neuron through a synapse. The communication protocol between such connected neurons is based on a train of spikes—a stream of binary numbers ('0' and '1' corresponding to spike off and on, respectively). A postsynaptic neuron generates a spike train in response to spike trains from presynaptic neurons in that messages encoded as presynaptic spike trains (similar to the Morse code) are encoded as another spike train by the postsynaptic neuron. This encoding is simultaneously performed along entire synaptic pathways, thereby macroscopically realizing high-level functionalities such as recognition and future prediction. SNN may also be viewed as another type of hypothesis for deep learning as an alternative to artificial neural networks (ANNs). Yet, it markedly differs from ANN given its potential capability of learning defined in a physical time domain.

There exist different strategies to build SNNs using dedicated hardware within the neuromorphic framework. Analog very-large-scale integrated circuitry (VLSI) is an originally proposed strategy in which the building blocks, e.g. spiking neurons and compatible synaptic circuits, are implemented using analog circuits.[8-10] Yet, hard wiring all spiking neurons through fan-in and out synapses appears inefficient and poor at network reconfiguration. Alternatively, digital event-routing techniques, such as address-event representation[11], offer a solution to efficient and reconfigurable network operation.

Most of neuromorphic systems built using analog building blocks are indeed based on mixed circuit.[1,3,12]

The digital neuromorphic prototypes, TrueNorth[2], SpiNNaker[4], and Loihi[5], highlight a fully digital circuit-based strategy in terms of flexibility of network configuration as well as neuron model parameters and learning algorithms. The progress in digital circuit fabrication techniques underpins high-speed and low-power operation of such digital neuromorphic systems. Additionally, a field-programmable gate array (FPGA) is a handy and cost-effective platform for a digital neuromorphic system. To date, various architectures of spiking neurons[13,14] and neuromorphic system architecture for reconfigurable SNN[15] have been built on FPGAs.

An emerging strategy is to utilize nonvolatile memory devices, e.g. oxide-based resistive memory, phase-change memory, magnetic tunnel junction, and floating-gate transistor, as synaptic devices.[16,17] Given needs for an enormous number of synapses in a neuromorphic system, replacing even in part mainstream static random access memory or content-addressable memory by such nonvolatile memories can remarkably enhance the areal density of synapses. Additionally, several nonvolatile memories represent multinary states that can further boost the areal density of synapses.[18]

Such cutting edge neuromorphic hardware can leverage its capability by the aid of a user-friendly complier, for instance, equipped with graphical user interface (GUI). An example is Nengo[19], a GUI-based compiler that readily builds an SNN on neuromorphic hardware. Lately, the complier has successfully been applied to Loihi and Braindrop.

SNNs largely vary in model complexity. The complexity is generally a measure of fidelity to biological neural network. That is, the more biologically plausible, the more likely the SNN needs high model complexity that consequently needs high computing power when simulated and large circuit footprint when realized in hardware. To date, available are a various neuron models with different degrees of fidelity to biological neurons, ranging from the simplest point neuron model, e.g. Stein's model[20], to the Hodgkin-Huxley model[21] with multiple compartments. Synapse models including its plasticity are even more diverse. It is aware that more than 100 types of neurotransmitters are involved in synaptic transmission (signal transmission between neighboring neurons through a chemical synapse) over the CNS.[22-24] Also, synaptic plasticity is known to incorporate a calcium signaling cascade that involves a number of different enzymes and their reactions along a number of reaction pathways.[22-25] This tremendous complexity in SNN may cause the difficulty in putting the SNN in a nutshell and approximating it to a simple form.

Simplifying the SNN but keeping the core functions importantly causes high areal density of neurons and synapses and thus low fabrication cost. Therefore, SNNs in neuromorphic engineering unnecessarily duplicate the CNS to understand biological neural processing. Instead, some "seemingly" important concepts in biological neural processing are taken and recreated in the SNN to realize desired functions such as inference and prediction as a consequence of training. We cannot elaborate all examples of such activities in this tutorial. Yet, we can categorize them as a few classes depending on (i) neural code in use, (ii) training scheme (algorithm), and (iii) functionality. Regarding the last two attributes, as the title of tutorial suggests, we will address *temporal learning* and its algorithm.

Note that the definition of temporal learning is unclear, the terminology is popularly invoked though. In this tutorial, temporal learning refers to *ad hoc* training of SNN with given time-varying input data in order to predict future reward correlated with the input data. In fact, such future reward prediction is the heart of reinforcement learning, and *ad hoc* training can be realized by the celebrated temporal difference (TD) learning.[26] The implementation of TD learning in SNN in association with neurophysiological learning rules, e.g. spike timing-dependent plasticity, is of concern in this tutorial. In this way, physically meaningless algorithmic time in general TD learning can be recreated in a physical time domain.

This tutorial concerns temporal learning of neuromorphic SNN from the scratch as follows. Section II is dedicated to basics of SNN including its general role in learning (Sub-section A), types of spiking neuron models (Sub-section B), and neuronal noise (Sub-section C). Subsequently, Section III addresses different neural codes in view of neurons as information encoders. Overviewed are spike-count code (Sub-section A), rate code (Sub-section B), and temporal code (Sub-section C). Such neural coding schemes may be suited for different synaptic modification rules. Introduction to such rules with neurophysiological plausibility is the main concern of Section IV, particularly, with regard to their relevance to TD learning. To this end, Section IV is dedicated to strategies for implementing a TD learning algorithm in SNN (Sub-section A), synaptic modification rules (Sub-section B), and exemplary linkages between TD learning and synaptic modification rule (Sub-section C).

**II. Basic of spiking neural network**

**A. Spiking neural network as a time-dependent hypothesis**

It is helpful to compare SNNs with ANNs that are much simpler than SNNs but nevertheless share many attributes with SNNs. Let us begin with feed-forward ANNs without recurrent connections. Feed-forward ANNs are directed graphs made of nodes and directed edges. Such a node has many names such as a unit, activation function, neuron, and so forth. Here we name it an "activation function" to distinguish it from "neuron" in a SNN. The activation function is a nonlinear function of the sum of weighted inputs from the preceding layer. Examples are simple McCulloch-Pitts model (binary model), logistic function, rectified linear unit, and so forth. Each activation function differently encodes the input following its mathematical form. For instance, the McCulloch-Pitts model encodes the input as a binary value (0 or 1) such that it is encoded as 1 if larger than a threshold, and 0, otherwise. The edges are given weight values that are important model parameters. The beauty is that this graph is a versatile mathematical model that can be transformed to desired mathematical functions $f$ by modifying the model parameters such as a weight matrix $\boldsymbol{w}$ and bias array $\boldsymbol{b}$. Formally, $f$ can be expressed as $f(\boldsymbol{x}; \boldsymbol{w}, \boldsymbol{b})$ where $\boldsymbol{x}$ denotes an input array.

The procedure of model parameter modification in line with purpose is referred to as training. Yet, we do not know (to be precise, we do not need to know) the exact form of the desired function that we are eventually given as a result of training. We cannot therefore directly modify the parameters. Instead, another function that guides the change of each model parameter is introduced—given by a function of the difference between the desired and actual outputs that can readily be measured. This function is popularly termed a cost function. The parameter modification proceeds from the output to the input layer, and thus this modification algorithm is referred to as "backpropagation".[27]

Notably, the ANN is a time-independent function since the model does not incorporate any time-varying functions.

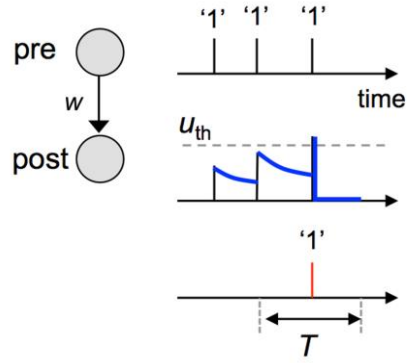

**Figure 1**. Pair of pre and postsynaptic neurons connected through a synapse whose weight is *w*. The upper panel shows a schematic of presynaptic spikes. The consequent evolution of postsynaptic membrane potential is illustrated in the middle panel. Stein's model[20] was used to evaluation the potential. When the potential reaches a threshold for spiking ($u_{th}$), a postsynaptic spike is evoked (lower panel).

SNNs are also directed graphs made of nodes (spiking neurons) and directed edges (synapses). Similar to the ANN, the SNN is a versatile mathematical model. Yet, unlike the activation function in the ANN, the spiking neuron does not directly take the sum of weighted inputs as a neural response-determining factor. Instead, the neuron involves a time-dependent state variable(s), e.g. membrane potential, that is a function of incident presynaptic spikes (input). Different from real-valued inputs in the ANN, a train of spikes as an input is a stream of binary numbers ('0' and '1'). The state variable temporally integrates such an incident bit stream, and determines the neuronal response.

Figure 1 illustrates a simple leaky integrate-and-fire model (Stein's model[20]) where a single state variable *u* evolves in response to input spikes and the neuron fires a spike when the state variable reaches a threshold ($u_{th}$). Spikes are also known as action potentials (APs). The firing is followed by state variable reset to the ground value and refractory period in which the neuron is inactive. The detail of spiking neurons will be addressed in the following section. The time evolution of the state variable indicates a *memory effect* that essentially endows the SNN with time-dependent characteristics. That is, the SNN is a time-dependent mathematical function *s* that can be expressed as $s[x(t), t; w, b]$ where input $x$ is a sequence of spikes in a physical time domain.

Note that there exist several neural networks for sequence learning, for instance, recurrent neural network[28,29] and long short-term memory[30]. Yet, these networks do not map the sequence onto a physical time domain in that only the sequence of events, rather than the interval between neighboring events, matters. The spiking neuron however encodes an input sequence of spikes (in a physical time domain) as also an output sequence of spikes (also in a physical time domain).

Consequently, learning algorithms should differ for such SNNs from ANNs. The neural engineering framework (NEF) offers a unique solution to training SNNs, enabling the SNNs to represent a wide variety of mathematical functions in a time domain.[31] Alternatively, TD learning[26] is capable of *ad hoc* update (rather than statistical optimization) on synaptic weight values to let the SNN eligible to predict future events (temporal learning). The relevance of neurophysiological learning rules to TD learning has been revealed in part, there remain significant uncertainties to be clarified though. This issue will be set aside until Section IV.

Notably, training ANNs often results in negative weight values for some edges—such negative weight inhibits an activation function such that the negatively weighted input to the activation function reduces the sum of inputs. In SNNs, a similar role is played by inhibitory synapses that feed negative input to a postsynaptic neuron, inhibiting the activity of the neuron as its name indicates. The other type of synapse representing positive weight is referred to as excitatory. The rule of thumb, Dale's principle coined by Eccles et al.[32], indicates that the same presynaptic neuron cannot possess both excitatory and inhibitory synapses in that an inhibitory (excitatory) neuron can have only inhibitory (excitatory) synapses at its axon terminals. This principle is not applied to ANNs. Nevertheless, Dale's principle is left as a rule of thumb given some recent findings of neurons possessing both types of synapses.[33]

**B. Spiking neuron models**

Intensive studies for the last decades have enriched spiking neuron models that vary in the fidelity to biological neurons. Generally, the higher the fidelity, the larger the model complexity, and thus the larger the number of state variables ($n_{sv}$). In the neuromorphic framework, given the limit of die area, simple neuron models are beneficially chosen unless the simplicity limits the high-level functionalities. In this regard, simple neuron models that involve one or two state variables ($n_{sv}$=1 or 2) are of concern in this section: Stein's model ($n_{sv}$=1)[20], leaky integrate-and-fire (LIF) model with excitatory postsynaptic current (EPSC)[34,35], Izhikevich's model[36], FitzHugh-Nagumo (FHN) model[37,38] ($n_{sv}$=2). Particularly, Izhikevich's model highlights the detailed similarity of its spiking dynamics to biological neurons.[36]

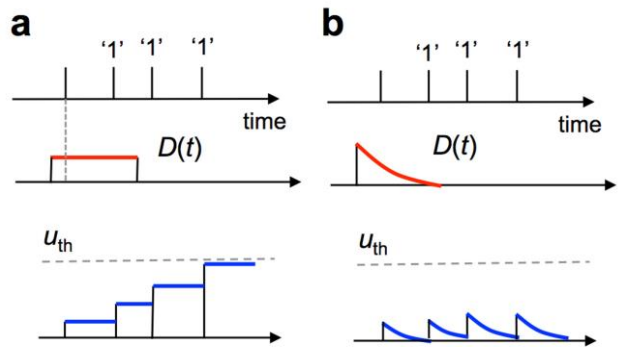

**Figure 2**. Integration of presynaptic spikes using a (**a**) constant window kernel and (**b**) exponentially decaying kernel and the resulting postsynaptic membrane potential evolution

**1. Neuron models involving a single state variable**

Available spiking neuron models are diverse. Yet, most of them involve a state variable(s) of a memory effect that essentially features dynamics neuronal behavior. The simplest model takes only one state variable ($n_{sv}$=1) that evolves amid incident presynaptic spikes, which is equivalent to somatic membrane potential dictating the spiking response. As illustrated in Fig. 1, the membrane potential abruptly increases upon the arrival of an input spike and subsequently decays away during the inter-spike interval (ISI). The state variable (membrane potential) is expressed as the following convolution:

$$u(t) = u_0 + a \int_0^t D(s) \cdot w \cdot \sigma(t-s) ds, \tag{1}$$

where $u_0$ and $a$ denote the base membrane potential and positive constant, respectively. $D(s)$ is a linear filter (kernel). $w$ means the synaptic weight. $\sigma$ is a sequence of $N$ input spikes ($i$th spike is arrived at $t_i$) through a synapse, which can be expressed as $\sigma(t) = \sum_{i=1}^{N} \delta(t - t_i)$. The synapse is given synaptic weight $w$.

Convolution for two different kernels is schematized in Fig. 2. The first kernel (Fig. 2a) is a constant window function between 0 and $t$. This constant kernel indicates no loss in memory so that it perfectly integrates the input spikes. The neuron models equipped with such constant kernels are referred to as integrate-and-fire (IF) models. In contrast, the second kernel (Fig. 2b) decreases in due course so that memory loss is obvious. Consequently, the kernel cannot perfectly integrate the input spikes due to the memory loss (leakage). The neuron models with such kernels are referred to as LIF models.

Particularly, the LIF model illustrated in Fig. 2b is called Stein's model that features an immediate rise in membrane potential upon the arrival of an input spike and consequent exponential decay in the potential.[20] TrueNorth employs a similar type of neuron model but the kernel simplified to a monotonic (rather than exponential) decay for an easier digital implementation.[2] The kernels for Stein's model and TrueNorth are

$$D_S(t) = H(t)e^{-t/\tau_u}, \tag{2}$$

and

$$D_{TN}(t) = max(-at + b, 0), \tag{3}$$

respectively. $H(t)$ and $\tau_u$ in Eq. (2) mean the Heaviside step function and a time constant for the decay. $a$ and $b$ in Eq. (3) are positive constant values that endows the kernel with a monotonic decay with a slope $a$ from $b$ in the first place. max choosing the larger value between the two components does not allow the kernel to fall below zero.

For both IF and LIF models, a spike is elicited when membrane potential $u$ reaches the threshold $u_{th}$, followed by resetting the membrane potential to its ground potential. Frequently, a refractory period for a few milliseconds is given after the membrane potential reset.

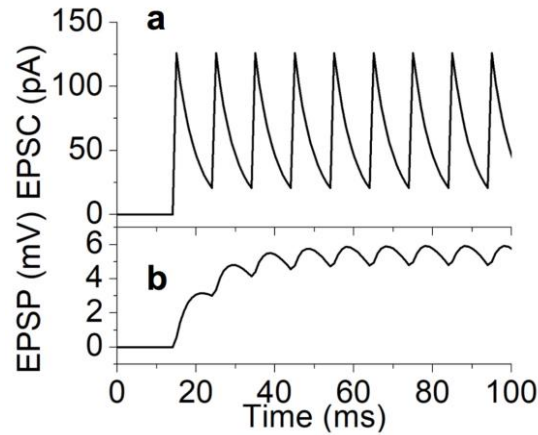

**Figure 3**. Evolution of (**a**) EPSC in response to periodical incident spikes and (**b**) the consequent EPSP calculated using Eq. (3).

## 2. Neuron models involving two state variables

Regarding neurophysiological synaptic transmission from an excitatory presynaptic neuron, excitatory neurotransmitters, e.g. glutamate, are released from the axon terminal of the excitatory presynaptic neuron. Subsequently, the released neurotransmitters largely increase the conductance of ion channels permeable to monovalent cations, e.g. $Na^+$ and $K^+$.[35] Such an influx of monovalent cations (mostly $Na^+$ ions) toward the intracellular is the substrate of EPSC. EPSC is a direct cause of a change in a potential difference across the membrane that integrates the EPSC. In view of the neuronal membrane integrating EPSC, it is conceivable to regard Eq. (1) as the evolution of EPSC (rather than membrane potential) and leaky-integrate this EPSC over time to produce the membrane potential. The model to this end is one step closer to the neurophysiological counterpart at the expense of one additional state variable, i.e. EPSC (*i*). EPSC *i* can be expressed as Eq. (1) after the replacement of $u$ by $i$. The EPSC is leaky-integrated over time using the following equation, yielding the response of membrane potential $u$ to the time-varying EPSC;

$$\frac{du(t)}{dt} = -\frac{u(t)}{\tau_u} + \alpha i(t). \tag{4}$$

Here $\alpha$ is a positive constant. Spike-firing and resetting the membrane potential processes are identical to the models in the previous section. That is, this model bases its dynamics on two state variables, $u$ and $i$. Membrane potential evolution in this LIF model amid incident presynaptic spikes is exemplified in Fig. 3. Loihi utilizes this type of neuron model.[5] The advantage of this model is that Eq. (3) is a linear differential equation without any higher-order term, rendering it simple to implement this model in both digital and analog neuromorphic systems.

Izhikevich's model leverages the capability of mimicking neurophysiological spiking behaviors of various modes.[36] Surprisingly, this model bases neurophysiological plausibility on only two variables ($u$ and auxiliary variable $v$) involved in the following two differential equations:

$$\frac{du(t)}{dt} = 0.04u(t)^2 + 5u(t) + 140 - v(t) + i(t), \tag{5}$$

and

$$\frac{dv(t)}{dt} = a[bu(t) - v(t)].$$

Here $a$ and $b$ are constant. When membrane potential $u$ reaches the firing threshold $u_{th}$, $u$ and $v$ are reset to $c$ and $v+d$, respectively. The beauty of Izhikevich's model is that a wide variety of spiking behaviors in detail can be manipulated by simply tweaking the four parameters $a$, $b$, $c$, and $d$. However, Eq. (5) is a nonlinear differential equation that includes a second-order term $u^2$. This may cause complexity in implementation in neuromorphic systems.

FHN model[37,38] also describes spike-firing dynamics using two state variables incorporated in two differential equations. A distinct feature of this model in comparison with the previous models is that no discrete changes in the state variables are necessary upon spike firing. Instead, the spiking behavior corresponds to continuously oscillating membrane potential. In digital implementation, a comparator to produce a spike in comparison with the spiking threshold is therefore unnecessary. Yet, this model also includes a nonlinear differential equation as for Izhikevich's model.[36]

Implementing firing rate adaption and refractoriness enriches neuronal dynamics, raising the neurophysiological plausibility of the model. Yet, it minimally costs one variable each. A common method for firing rate adaption implementation is to introduce an additional state variable similar to Eq. (1), which accelerates the membrane leakage.[35] Thus, spiking events are suppressed to some extent depending on the contribution of the accelerated leakage to the integration. Likewise, refractoriness minimally costs one variable (time stamp) that is a record of a last spike timing in case of digital neurons. Analog neurons need the circuitry generating a particular window function whose width is the refractory period.

### 3. Multi-compartment models

All these models quantitatively describe the dynamics of somatic membrane potential. In a point neuron model, the neuron is given only one integrator that integrates the input (presynaptic) spikes over all dendritic spines. In contrast, multi-compartment models distinguish dendritic membrane potential from somatic membrane potential in that each dendritic spine is given an independent integrator that evaluates dendritic potential in response to the input spikes into the dendritic spine. The soma integrates all input spikes over all dendritic spines to evaluate somatic potential. Yet, given the physical distance between a dendritic spine and the soma and consequent attenuation of a spike along the distance, a raise in the somatic potential upon an input spike is less than the dendritic potential. Generally, the soma is considered to be the only compartment able to produce spikes. Such multi-compartment models support more biologically plausible learning models involving spikes backpropagating from soma to dendritic spines. Backpropagating spikes following the onset of a somatic spike affect the dendritic potential of each dendritic spine to the extent that differs for different dendritic spines (in location) of different dendritic potential.[39] Therefore, such learning models need a state variable for each spine to support local learning models, and dendritic potential is a suitable state variable.

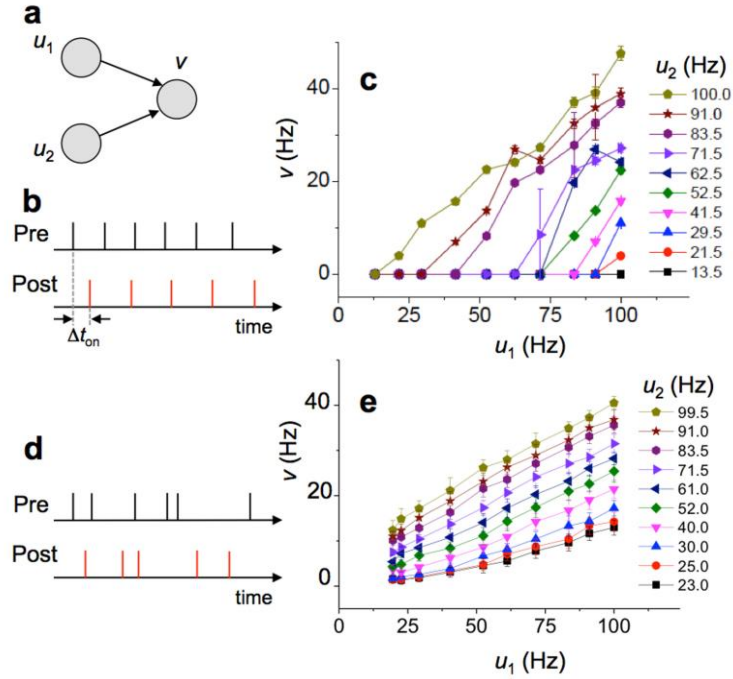

**Figure 4**. (**a**) Schematic of two presynaptic neurons sharing a single postsynaptic neuron. $u_1$, $u_2$, and $v$ denote the firing rates of the two presynaptic neuron, and postsynaptic neuron, respectively. The postsynaptic firing rate was evaluated taking into account (**b**) a random relative onset of a postsynaptic spike to a first presynaptic spike. Both pre and postsynaptic spikes were assumed to be perfectly periodic. The calculated postsynaptic firing rate on average on 50 trials was plotted with $u_1$ and $u_2$ in (**c**). (**d**) A schematic of Poisson pre and postsynaptic spikes and (**e**) calculation results of average $v$ with $u_1$ and $u_2$.

**C. Noise in biological neurons and its implementation in spiking neurons**

An important attribute of neural coding in SNNs is random noise (variability). In biological neurons, a random fluctuation in membrane potential is noticed. Such a fluctuation partly arises from stochastic behavior of ion channels in which the voltage-gated ion channels are open and closed in a stochastic fashion, resulting in a stochastic transfer of ions through the channels.[40-42] Another cause of the random fluctuation is stochastic synaptic transmission causing an uncontrollable fluctuation in EPSC that is integrated on the membrane resulting in membrane potential.[42,43] Additionally, the same holds for inhibitory synapse in that the consequent fluctuation in inhibitory postsynaptic current (IPSC) significantly perturbs the membrane potential of the postsynaptic neuron.[44,45] The network effect due in part to crosstalk between neighboring neurons also causes a random fluctuation in membrane potential.[46,47] That is, neighboring neurons (closely sharing the extracellular medium) may affect each other via the extracellular medium whose potential can be perturbed upon spike generation and propagation in one neuron.[46,47]

Altogether, such stochastic effects cause a variability in spiking, which is readily parameterized by variability in ISI. The variability in ISI indicates the breakdown of perfect periodicity in spikes in a time domain even under a time-independent external input stimulus. A probability distribution function (PDF) fitted to a measured statistics on ISIs frequently reveals the nature of stochasticity. Often, the statistics supports a gamma distribution function, different distributions varies in the function parameters though.[48] Particularly, the PDF in detail markedly differs for different ratios of inhibitory to excitatory synapses within the SNN by balanced excitation and inhibition.[45] Such a ratio thus offers a means of tweaking an ISI PDF when designing an SNN.

A question as to if we should implement such variability in ISI in neuromorphic systems arises for the moment. In the digital computing framework, variability should be avoided because it disturbs perfect periodicity and simultaneous operation in synch with a system clock. However, asynchronous SNNs (without system clock) perhaps make active use of such variability. Let us consider a postsynaptic neuron shared by two presynaptic neurons as shown in Fig. 4a. Assuming invariant ISI of output spikes from both presynaptic neurons, the response of the postsynaptic neuron is mainly dictated by the onset of an input spike train into the postsynaptic neuron from each presynaptic neuron. Because of the lack of system clock in the asynchronous SNN, the onsets are uncontrollable, and the response of the postsynaptic neuron significantly varies upon trials of random onsets (see Figs. 4b and 4c). In contrast, the response of the postsynaptic neuron to two trains of Poisson spikes exhibits remarkably small variations (Figs. 4d and 4e). This example may ascertain the importance of random variability in asynchronous SNNs within the rate-coding framework.

Poisson spikes are generated using a Poisson process—a renewal process that has the same probability of event occurrence in all time bins of concern. The resulting intervals between neighboring events follow an exponential distribution over the interval.[35] In a digital system, a usual procedure of Poisson spike generation at particular firing rate $r$ in time period $T$ is as follows:

(i)          dividing the time period $T$ into $N$ bins (each bin size $\Delta t$ is $T/N$).

(ii) generating an independent and identically distributed random number *rand* (0 ≤ *rand* ≤ 1).

(iii) giving a spike to the first time bin if *rand* < *r*Δ*t*, no spike otherwise.

(iv) repeating (ii) and (iii) up to the last (*N*th) time bin.

A PDF of the ISI between spikes generated using this method (*r*=50, Δ*t*=1ms, *T*=100) is displayed in Fig. 5a. Given a presynaptic (Poisson) neuron (Pre) in the inset, the postsynaptic neuron (Post) sees incident Poisson spikes.

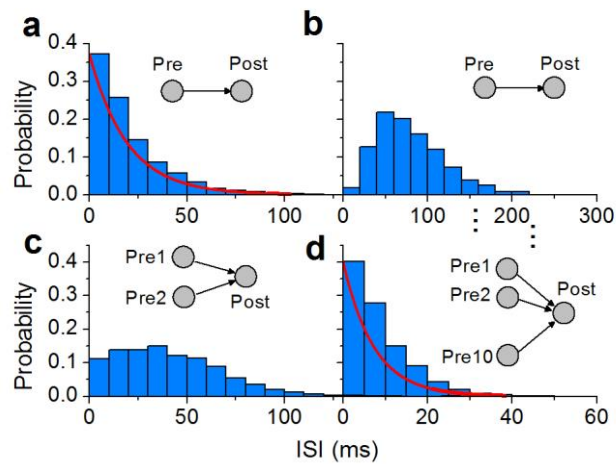

**Figure 5**. (**a**) ISI PDF of a Poisson neuron (termed Pre), conforming to the exponential distribution (*r*=50). The postsynaptic neuron in the inset (termed Post) is therefore subject to the Poisson spikes. (**b**) The PDF conforming to the gamma distribution. The data were produced for *r*=50 and *n*=4. As such, the postsynaptic neuron is subject to the spikes whose ISI follows the gamma distribution. (**c**) A PDF of the interval between nearest neighboring spikes arriving at the postsynaptic neuron (Post) when two presynaptic neurons (Pre1 and Pre2) simultaneously emit spikes. Each presynaptic neuron, however, emits spikes whose ISI conforms the gamma distribution (*r*=50 and *n*=4). (**d**) In the case of ten presynaptic neurons (Pre1 – Pre10) that simultaneously fire spikes that are directed to a single postsynaptic neuron (Post).

Generating more realistic spikes (whose ISIs follow a gamma distribution) needs to involve a more complexity than the abovementioned Poisson spikes. Yet, a simple way to generate spikes conforming to the gamma distribution is to remove all Poisson spikes (generated using the above method) except every *n*th spike.[49] Figure 5b is an example of ISI distribution of a spike train (*n*=4) generated using this method ($r$=50, $\Delta t$=1ms, $T$=100). The postsynaptic neuron (Post) in the inset is consequently stimulated by the spikes following the gamma distribution. Nevertheless, Poisson spikes are a sensible approximation to the more realistic spikes. When spikes from different neurons (each of which generates spikes following a gamma distribution) are merged, the ISI distribution is known to follow an exponential PDF, which is evident in Figs. 5c and 5d.

**III. Neural coding schemes**

The directed edges as synapses in SNNs capture the biological unidirectional synaptic transmission through a chemical synapse from a presynaptic to a postsynaptic neuron. Therefore, every synapse is given a pair of presynaptic and postsynaptic neurons. A postsynaptic neuron fires spikes (to be precise, postsynaptic spikes) to presynaptic spikes. Such postsynaptic spikes then work as presynaptic spikes for a subsequent synapse and elicit postsynaptic spikes from the postsynaptic neuron. This process is repeated along the directed edges in a given SNN. When viewing a neuron as a neural information coder, this neurophysiological behavior is a means of conveying information along the directed edges. For instance, sensory neurons encode different external stimuli as distinct spiking features so that, given a particular spiking feature of a sensory neuron, the external stimulus can be estimated. To this end, the input and output spikes in a time domain should be described in terms of particular physical quantity (representing the spiking feature) that is coded by the neuron.

As a matter of fact, neural coding is not as simple as that for ANNs where the real-valued sum of weighted inputs is coded as a real-valued output according to a given activation function. We introduce several examples of neural codes with reference to biological SNN.

Mutual information is frequently used as a measure of the information transferred by a neuron (or population).[50-52] A neuron as a neural information coder transforms an input signal (*In*) to an output signal (*Out*), and its reliability is evaluated in terms of the amount of information obtained about the input signal by seeing the output signal (mutual information). The mutual information $I(In;Out)$ is

$$I(In; Out) = H(In) - H(In|Out), \qquad (6)$$

where *H*(*In*) is the information of input, and *H*(*In*|*Out*) is the conditional information of input given particular *Out* signals. For instance, a sensory neuron that is orientation-selective selectively response to different orientations (encoding), and the information conveyed by this neuron is a measure of the extent to which a given input is correctly estimated from the paired output (decoding), which is evaluated by *H*(*In*|*Out*) in Eq. (6).

Practically, the mutual information in Eq. (6) is dictated by the decoding capability *H*(*In*|*Out*) because of uncontrollable variability (e.g., ISI) of neuronal response to input. In an extreme case of perfect decoding (perfect estimation), *H*(*In*|*Out*) falls to zero the maximum information *H*(*In*) is conveyed without loss. In the opposite extreme case, the variability is so significant that the correlation between *In* and *Out* vanishes, and thus *In* and *Out* become independent of each other. This case sets *H*(*In*|*Out*) to *H*(*In*), leading to an *I*(*In*;*Out*) of zero according to Eq. (6). The three coding schemes addressed in this section represent *Out* using different physical quantities that consequently vary in *H*(*In*|*Out*) given the same *In* and variability. That is, the degree of information loss due to the variability differs for different schemes.

**A. Spike-count code**

When a neuron is given a perfect integrator (as for the kernel in Fig. 2a), the neuron counts a number of input spikes until firing. The integrator excludes a leakage contribution to the membrane potential so that the neuronal response is solely dictated by the number of input spikes for given synaptic weight irrespective of how fast the spikes arrive at the postsynaptic neuron (i.e. input firing rate) and when the spikes arrive (i.e. input spike-timing). The spike-count code for a sensory neuron expresses *Out* in Eq. (6) using the number of spikes such that different input stimuli induce different spike numbers, endowing the neuron with input-selectivity. An advantage is that this scheme is most robust to variability in ISI among the three schemes because the ISI is excluded in the representation of neuronal information. A schematic of spike-count code is displayed in Figs. 6a and 6c, for a sensory neuron and a pair of presynaptic and postsynaptic neurons, respectively. Figure 6c depicts postsynaptic spikes in response to a presynaptic spike train for two different weight values ($w_1 < w_2$). The postsynaptic neuron fires a spike to a certain number of incident presynaptic spikes.

It is noticed that rate code is at times also referred to as spike-count code given that firing rate is equivalent to spike counts within a time window.[53] However, the spike-count in this tutorial is not confined within such a time window.

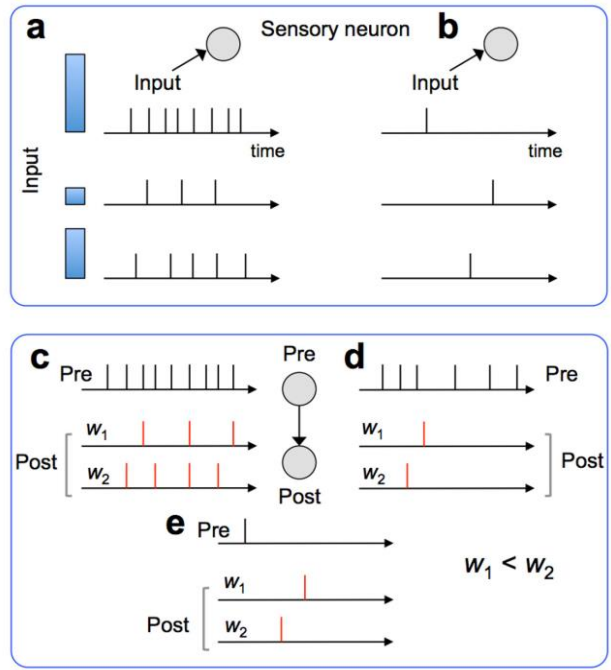

**Figure 6**. Illustrations of (**a**) spike-count and rate coding and (**b**) latency coding concepts for sensory neurons. The bar on the left side of each panel indicates the strength of input into a sensory neuron. The stronger the input, the earlier the onset of an induced spike. Encoding a presynaptic spike train as a postsynaptic spike train for (**c**) spike-count code and (**d**) rate code at two different weight values ($w_1 < w_2$). Spike-count code elucidates a presynaptic spike after certain number of incident presynaptic spikes as in (**c**). Rate code however elucidates a presynaptic spike if the presynaptic spike rate is sufficiently large to raise the postsynaptic membrane potential above the firing threshold. (**e**) Latency code accounts for different firing latencies for different weight values.

**B. Rate code**

Rate code represents *Out* (in Eq. (6)) of a sensory neuron by means of firing rate. Firing rate is a measure of the number of spikes in a certain time window *T* (see Figure 1), and thus refers to the average spike number per unit time. Figures 6a and 6d schematize a postsynaptic spike train in response to a sensory input and a presynaptic spike train, respectively. Different from spike-count code, rate code allows postsynaptic spiking when a presynaptic spiking rate is sufficiently high to elevate the postsynaptic membrane potential with leakage above the firing threshold. In Fig. 6d, the first three incident presynaptic spikes have a high firing rate so that a postsynaptic spike is elicited whereas the last three presynaptic spikes are of a low rate, and thus inducing no postsynaptic spike.

It should be noted that rate code is often called spike-count code as opposed to temporal code because rate coding essentially count spikes to evaluate the temporal-average of incident spikes.[35] However, we distinguish rate code from spike-count code given their distinct coding schemes with distinct types of integrators. In neuroscience, firing rate has long been a standard measure of neuronal activity and handy variable in neural codes.[54] Firing rate can be given by a continuous function of time in contrast to spike timing-based description that is discrete in time. It therefore offers analytical solutions to the response of individual neurons and neuronal populations, and synaptic plasticity models in theoretical neuroscience.[35] The leaky integrator is the substrate of rate coding in that the leakage contribution differentiates among spike trains of the same number of spikes but different rates. That is, the larger the input firing rate, the larger membrane potential is achieved. Note that, given the leaky integrator in a spiking neuron, the neuron is naturally suited for rate coding.

A disadvantage is that rate coding is essentially time consuming because each neuron should integrate input spikes sufficient to evaluate the temporal average of spike number per unit time. There are a few workarounds for this issue; (i) a rise in firing rate as a whole over the entire network, which shortens the integration time and (ii) population representation in which a group of neurons fire simultaneously so that a postsynaptic neuron shared among the population receives sufficient input spikes in a short time window, each neuron in the population keeps a low firing rate though. The latter leaves the important premise that the spike from each neuron in the population is independent of each other, and thus no temporal correlation is established.

Note that because firing rate is a temporal average value, it averages out variability in ISI to some extent. This advantageous attribute of rate coding supports its robustness to variability in ISI also the high reliability of temporal averaging due to the variability as explained in Section II.

**C. Temporal code**

Temporal code—representing *Out* (in Eq. (6)) using spike timing—brings up an objection to rate code in that rate coding is not sufficiently fast to duplicate the fast decision-making of biological brains. As such, evaluating temporal-average takes some time, which is inconsistent with biological observations. Also, researchers advocating temporal code indicate that decision-making neurons perform actions after a few input spikes[55,56] as opposed to the acceleration of temporal-averaging by means of population representation. Alternatively, temporal code bases neural information on individual spikes rather than their temporal average. A few spikes convey important neural information along neurons through directed edges. A conceivable way to this end is to take spike timing as a code. The latency coding scheme proposed by Thorpe is a good example of temporal coding.[56] Figure 6b illustrates an example of latency coding for a sensory neuron in which the latency (*Out*) differs for different inputs (*In*). The input intensity dictates the spike timing in that the stronger the input the sooner a spike is evoked (the lower the spiking latency). This makes good sense regarding the LIF behavior of a neuron.

The important premise is that spikes elicited from neurons should be sparse in time to avoid any temporal-average effects. Ideally, a single spike (its spike timing) is sufficient to convey the input information as shown in Figs. 6b (sensory neuron) and 6e (a pair of presynaptic and postsynaptic neuron). Thus, a threshold for spiking should be adjusted to be reached upon a single spike or a few spikes. This way deprives the leaky integrator of the natural temporal-averaging effect that makes the latency information of no use.

It should be noted that temporal code is prone to error when the variability in ISI is present because the variability directly corrupts the *Out* signal dissimilar to the rate code taking temporal-average quantity as *Out*.

**IV. Temporal learning of spiking neural network**

As such, temporal learning enables SNN to be eligible to predict future reward for time-varying input data. Here future is defined in a physical time (rather than algorithmic time) domain due to the fact that the spiking neuronal behavior involves physical time. Also, *ad hoc* training (rather than batch training) is desirable. To this end, TD learning[26] appears suitable with regard to its *ad hoc* update feature for future reward prediction. The original TD learning has little relevance to SNN.[57] The first step is therefore "coarse" implementation of TD learning in SNN. Enriched synaptic modification rules may hint at desirable refinement strategies. The compatibility such local synaptic modification rules with TD learning therefore needs to be looked into beforehand. This as reverse engineering may also shed light on understanding of the biological brain such that the roles of synaptic modification in high-level brain functionalities are indirectly deduced.

The neurophysiologically plausible learning rules introduced below address synaptic modification upon the states of pre and postsynaptic neurons. To be specific, their activities (temporal-average firing rates) or spike timings determine not only the degree of weight modification also the direction of modification (potentiation or depression). Such a local rule applied to all pairs of pre and postsynaptic neurons in a given SNN in parallel drives the whole network towards a certain configuration that supports a desired high-level function such as prediction.

**A. Temporal difference learning in spiking neural network**

**1. General temporal difference learning**

Assume a time-varying input $u(t)$, the consequent reward $r(t)$, and prediction $v(t)$, all in the time period $(0 - T)$. Prediction $v(t)$ is assumed to be a function of not only $u(t)$ but also $u$'s in the previous time steps, i.e. $u(t\text{-}1)$, $u(t\text{-}2)$, $u(t\text{-}3)$, ..., but their contributions differ. That is, the past inputs influence the current prediction, and their contributions are quantified by weight $w(i)$, where $i = 0, 1, ..., t$. $w(i)$ quantifies the influence of the input $i$th time steps before $t$, i.e. $u(t - i)$, on $v(t)$ that can therefore be expressed as

$$v(t) = w(t)u(0) + w(t-1)u(1) + \cdots + w(0)u(t) = \sum_{s=0}^{t} w(t-s)u(s). \tag{7}$$

TD learning regards $v(t)$ as the prediction of weighted sum of future rewards at the current time step $t$, $[r(t)]$, and all succeeding time steps until $T$, $[r(t) + r(t+1) + \ldots + r(T)]$.[26] The weight for each term is given by a discounting factor $\gamma$ ($\leq 1$). Its target (correct) value $v'(t)$ is therefore

$$v'(t) = \sum_{s=0}^{T-t} \gamma^s r(t+s). \tag{8}$$

$v(t)$ is different from $v'(t)$ before learning, and $w$'s in Eq. (7) are optimized such that the resulting $v(t)$ asymptotically approaches $v'(t)$. The temporal model in Eq. (7) is identical to a spatial model of $(t+1)$ parallel nodes and a node shared among the parallel nodes as depicted in Fig. 7. This linear neural network can readily be trained using the delta learning rule elaborated in Appendix. The weight array is therefore updated at time step $t$ such that

$$w(i) \rightarrow w(i) + \eta \delta(t) u(t-i) \quad \text{for all } i\text{'s } (\leq t), \tag{9}$$

where $\delta(t) = \sum_{s=0}^{T-t} \gamma^s r(t+s) - v(t)$. $\eta$ is a learning rate. Now the difficulty in training is that evaluating $\delta$ needs the entire future rewards at all time steps until $T$, rendering it impossible to update all weight values at every time (i.e. *ad hoc* update). Instead, the weight update is postponed until the end of the training period $T$ at which the full reward information at every time is eventually unveiled.

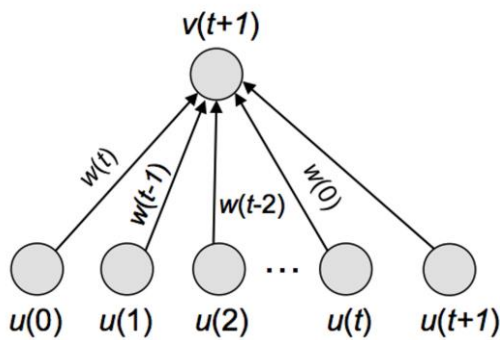

**Figure 7**. Graphical description of TD learning.

The beauty of TD learning lies in the fact that the sum of future rewards in Eq. (8) can recursively be approximated to

$$v'(t) = \sum_{s=0}^{T-t} \gamma^s r(t+s) = r(t) + \sum_{s=0}^{T-t-1} \gamma^{s+1} r(t+1+s) = r(t) + \gamma v(t+1).$$

Thus, $\delta$ can be re-written as $\delta = r(t) + \gamma v(t+1) - v(t)$. The weight update rule is consequently given by

$$w(i) \rightarrow w(i) + \eta[r(t) + \gamma v(t+1) - v(t)]u(t-i) \quad \text{for all } i\text{'s } (\leq t), \tag{10}$$

Note that the training is one time-step delayed in that $v(t+1)$ should be acquired to update $w(0)$ as for $i = 0$ in Eq. (10).

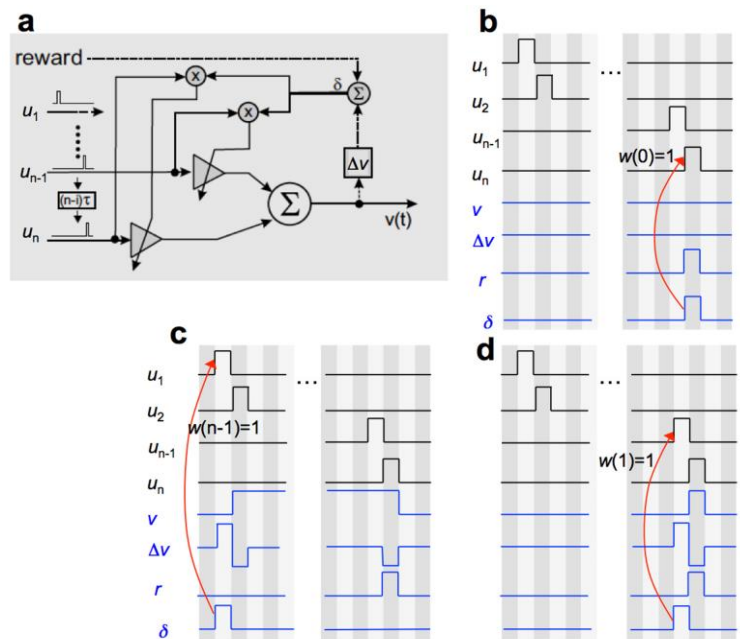

**Figure 8**. (**a**) Circuit diagram of TD learning algorithm implemented in SNN. Reproduced with permission.[57] Copyright 2005, MIT Press. (**b**) A timing diagram of the TD learning algorithm implementation for the (**b**) first, (**c**) second, and (**d**) last epoch. The red arrows indicate weight update given nonzero input $u$ and $\delta$.

## 2. Temporal difference learning implementation in spiking neural network

Original TD learning did not explicitly choose SNN as a network model. It took a while for a feasible implementation of TD learning in SNN to appear.[58] An intuitive way to do so is to replace the input $u$ and output $v$ in Eq. (10) by spikes in which $u$ and $v$ at a given time are endowed with either '1' (spiking) or '0' (non-spiking) because of the all-or-nothing feature of spikes. An exemplary algorithm is shown in Fig. 8a, which was proposed by Montague et al.[58] This simple implementation aims to establish a causal correlation between the signal from the input neuron ($u$) and delayed reward ($r$). The established causality after full training lets a single spike from the input neuron elicit a persisted '1' signal until the delayed reward is eventually given. Recall that in the TD learning the output signal indicates reward prediction as for Eq. (8). The '1' signal maintained until the actual reward is given therefore corresponds to a predicted reward for the input signal. Additionally, it makes clear sense that the output '1' signal vanishes upon the delayed reward is eventually given because prediction is no longer valid insomuch that it comes true.

The SNN in Fig. 8a needs $n$-1 auxiliary neurons other than the input and output neurons if the reward spike is given at the $n$th time step. Note that the discount factor $\gamma$ was set to one. The auxiliary neurons relays nonzero TD term [$v(t+1) - v(t)$], initially only at the $n$th time step, from the $n$th time step eventually to the first time step where the input appears such that one time step backwards each training epoch. The circuit diagram realizing the TD learning in SNN is illustrated in Fig. 8. The timing diagram for the first two and last epochs is depicted in Figs. 8b, c, and d, respectively. Note that $\Delta v$ in this figure is the TD term that is time-shifted forward by one time step. A nice review on this topic can be seen in a paper by Wörgötter and Porr.[57]

## B. Neurophysiological synaptic modification models

The Hebb rule is the substrate of synaptic modification based on the correlation between the pre and postsynaptic neurons.[59] The general term "Hebbian learning" covers correlation-based synaptic modification (precisely, potentiation) rules including the basic Hebb rule and its modifications, e.g. the Bienenstock-Cooper-Munro (BCM) rule[60,61] and Cooper-Liberman-Oja (CLO) rule[62]. In these rules, synaptic modification is a function of pre and postsynaptic firing rates (activities). The correlation between pre and postsynaptic neurons is unnecessarily expressed by firing rate (many spikes). Instead, single pre and postsynaptic spikes capture their correlation in terms of spike timing, and the consequent synaptic modification also belongs to the class of Hebbian learning. An example is spike timing-dependent plasticity (STDP).[63-65] However, the BCM rule, CLO rule, and STDP conditionally gives rise to depression as well, which is referred to as anti-Hebbian learning. This conditional bidirectional synaptic modification clearly distinguishes these models from the basic Hebb rule.

**1. Basic Hebb rule**

The basic Hebb rule provides a primitive way to adjust synaptic weight between a pair of neurons with regard to their correlation.[59] The celebrated statement "cells that fire together wire together" explains *correlation*-based weight enhancement, to be specific, long-term potentiation (LTP).[59] This rule involves two factors representing the activities of pre and postsynaptic neurons that sandwich a synapse subject to modification. Albeit too simple, the Hebb rule still offers the main framework of learning in biological SNNs. Given no explicit condition of weight modification upon temporal correlation between the firing cells, the basic Hebb rule describes a continuous change in weight ($w_{ji}$) between a pre and postsynaptic neurons whose temporal-average firing rates are $a_i$ and $a_j$, respectively, as follows

$$\tau_w \frac{dw_{ji}}{dt} = a_j a_i. \tag{11}$$

$\tau_w$ denotes a time constant of weight change. This equation is readily applied to train an SNN with time-dependent pre and postsynaptic activities.

Yet, this basic Hebb rule falls short of explaining several detailed attributes of synaptic modification. First, it cannot account for long-term depression (LTD) since the right-hand side of Eq. (11) does not fall below zero. It evident that LTD is an important gradient of successful learning together with LTP and that a synapse can experience both LTP and LTD depending on given conditions. In this regard, LTD is referred to as homosynaptic depression.[66] Therefore, LTD is classified as anti-Hebbian learning. Second, this simple model hardly captures a temporal correlation between pre and postsynaptic spikes because the model takes temporal-average quantities ($a_i$ and $a_j$) as inputs. Third, no upper limit of weight is set in the basic Hebb rule in that the weight in Eq. (11) unlimitedly increases as far as the pre and postsynaptic neurons are active (the right-hand side of Eq. (11) is positive). This is at odds with neurophysiological behavior of synaptic modification where weight saturation is evident. Irrespective of this neurophysiological fidelity, such unlimited weight growth causes network stability so that it should be avoided.[35]

## 2. Bienenstock-Cooper-Munro rule

The BCM rule empirically reconciles the homosynaptic depression (LTD) with LTP to different extent for different postsynaptic activities, and thus combining LTD and LTP using a single model.[60,61] To be specific, the BCM empirically elucidates that, in the presence of presynaptic spikes, the induced postsynaptic activity below (above) a threshold causes LTD (LTP) in line with the neurophysiological data.[61,66] In addition, this threshold is history-dependent (has a memory effect) in that the larger the previous postsynaptic activity on average, the larger the threshold that requires the larger postsynaptic activity for a further increase in synaptic weight.

The threshold offers the substrate of input-selectivity evolution laying the foundation for selective neuronal responses to different inputs.[60,61,67] LTP continues onwards at the synapses along a positive feedback loop such that the synaptic inputs whose sum evokes postsynaptic activity greater than the threshold cause the synapses to gain more weight. This supports an even larger postsynaptic activity resulting from the same synaptic inputs. In contrast, the synapses whose synaptic inputs are insufficient for the postsynaptic activity greater the threshold fall behind competition with the winning synapses, and thus the weight keeps decreasing along a negative feedback loop. The combination of the positive and negative feedback loops eventually distinguishes an ensemble (pattern) of synaptic inputs, which causes the higher postsynaptic activity, from the others—referred to as neuronal selectivity.

Note that the sliding threshold however hinders unlimited weight growth such that the increasing threshold with the postsynaptic activity is reconciled with the positive feedback (resulting in otherwise unlimited growth). This is a distinct feature of the BCM rule in comparison with the CLO rule[62] that also deploys a threshold separating LTD and LTP but constant irrespective of the history of the postsynaptic activity.

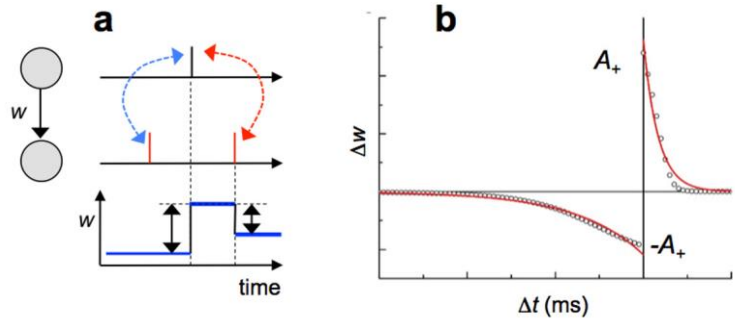

**Figure 9**. (**a**) Weight update using the basic STDP rule. The blue arrow indicates pre-post order leading to potentiation and the red arrow post-pre order leading to depression. (b) A schematic STDP behavior.

**3. Spike timing-dependent plasticity**

STDP elucidates synaptic modification determined by the temporal order of pre and postsynaptic spikes.[63-65] A postsynaptic spike following a presynaptic spike within a certain time-window (pre-post order) causes LTP while the opposite order (post-pre order) LTD as shown in Fig. 9. Notably, the post-pre temporal order realizes homosynaptic depression (anti-Hebbian learning). The simplest model empirically describing this intriguing observation (between presynaptic neuron $i$ and postsynaptic neuron $j$) is a spike-pair-based model as follows

$$\Delta w_{ji} = \begin{cases} A_+ e^{-(t_{post}-t_{pre})/\tau_+} & if\ t_{post} > t_{pre} \\ -A_- e^{-(t_{pre}-t_{post})/\tau_-} & if\ t_{pre} > t_{post} \end{cases}, \qquad (12)$$

where $t_{post}$ and $t_{pre}$ respectively denote timings of post and presynaptic spikes. The STDP behavior in detail is determined by $\tau_+$ ($\tau_-$) and $A_+$ ($A_-$) for LTP (LTD) as shown in Fig. 9b. This is an easy-to-implement model insomuch that it can readily be implemented in SNN hardware for on-chip learning, for instance, fully digital SNN (e.g. Loihi[5]) and analog synaptic circuits[68].

The implementation needs two state variables $m_{pre}$ and $m_{post}$ that evolve as follows:

$$m_{pre} = \begin{cases} A_+ & when\ t = t_{pre} \\ A_+ e^{-(t-t_{pre})/\tau_+} & when\ t > t_{pre} \end{cases}, \tag{13}$$

and

$$m_{post} = \begin{cases} A_- & when\ t = t_{post} \\ A_- e^{-(t-t_{post})/\tau_-} & when\ t > t_{post} \end{cases}. \tag{14}$$

These variables are kept track of through the implementation and read when a pre or postsynaptic spike is evoked in that $m_{pre}$ ($m_{post}$) is read and determines $\Delta w$ when a postsynaptic (presynaptic) spike is elicited. That is,

$$\Delta w_{ji} = \begin{cases} m_{pre}(t_{post}) & at\ t_{post} \\ -m_{pre}(t_{pre}) & at\ t_{pre} \end{cases},$$

identical to Eq. (12).

Accordingly, every pair of pre and postsynaptic spikes (unnecessarily pairs of pre and postsynaptic spikes in the closest temporal vicinity) can induce the STDP. $M$ presynaptic spikes and $N$ postsynaptic spikes allow total $M \times N$ pairs in a time domain so that all pairs should be considered to evaluate the final synaptic weight, a pair of pre and postsynaptic spikes separated far negligibly contributes to the total weight change though.

## 4. Paradox of the simple STDP rule and unified models

The simple STDP model can easily be mapped onto an activity domain as shown by Izhikevich and Desai.[69] In the simplest case, this mapping ignores any temporal correlation between pre and postsynaptic spikes such that they are regarded as Poisson spikes at temporal-average firing rates $a_i$ and $a_j$ for the pre and postsynaptic neurons, respectively. In this case, the expected weight change $\Delta w$ during a training period $T$ is expressed as

$$\Delta w_{ji} = (A_+ \tau_+ - A_- \tau_-) a_i a_j T,$$

which is equivalent to the following differential form,

$$\frac{dw_{ji}}{dt} = (A_+ \tau_+ - A_- \tau_-) a_i a_j. \tag{15}$$

The derivation is elaborated in Appendix. Notably, Eq. (15) is identical to Eq. (11) other than $(A_+\tau_+ - A_-\tau_-)^{-1}$ as a replacement for $\tau_w$ in Eq. (11). Given the biological observations, the LTD time-window dictated by $\tau_-$ is a few times as wide as that for LTP ($\tau_+$), and the difference between $A_+$ and $A_-$ is relatively smaller.[63,64] Thus, interpreting the neurophysiological STDP using this simple model yields that $A_+\tau_+ - A_-\tau_- < 0$, implying an unlimited decrease in weight with nonzero $a_i$ and $a_j$ as opposed to the Hebb rule. This is at odds with the neurophysiological data that clearly support the Hebb rule dominant over the STDP under high temporal-average firing rates.[65] If $A_+\tau_+ - A_-\tau_- > 0$, Eq. (15) becomes equal to Eq. (11) so that this model includes the aforementioned limitations (including the lack of LTD) of the Hebb rule.

One may conceive that a temporal correlation between pre and postsynaptic spikes (ignored in the above discussion) yields a different view of the model in an activity domain. However, Izhikevich and Desai has theoretically ascertained that such a temporal correlation barely fits the model to the neurophysiological data including the threshold for LTP.[69] A conceivable correlation arises from a causality between a pre and postsynaptic spike that is evoked by the presynaptic spike and thus succeeding it. Regarding this causality, the probability of postsynaptic spiking is non-uniform such that the probability is centered at $t_{\text{pre}} + \Delta t_{\text{post}}$ ($\Delta t_{\text{post}} > 0$) together with a constant background probability.

The detailed derivation is given in Appendix. The following $\Delta w_{ji}$ given the correlation holds:

$$\Delta w_{ji} = (A_+ \tau_+ - A_- \tau_- + l) a_i a_j T,$$

where $l$ is most likely to be a positive constant value. This equation is also written as a differential form as follows

$$\frac{dw_{ji}}{dt} = (A_+ \tau_+ - A_- \tau_- + l) a_i a_j. \tag{16}$$

$(A_+\tau_+ - A_-\tau_- + l)$ in this equation is positive when $l$ outweighs $(A_+\tau_+ - A_-\tau_-)$ that is most likely negative given the neurophysiological observations. Otherwise, $(A_+\tau_+ - A_-\tau_- + l)$ is negative. Yet, irrespective of its sign, such a temporal correlation barely supports the presence of the threshold and a transition from LTD to LTP with an increase in postsynaptic activity.[65,66]

This paradox may indicate that this simple phenomenological STDP rule is merely an attribute of the universal rule that is approximated to the simple STDP rule if and only if both pre and postsynaptic firing rates are sufficiently low (i.e. the ISI is sufficiently large) to avoid multiple reads of a presynaptic (postsynaptic) state variable by successive postsynaptic (presynaptic) spikes. In this regard, the STDP rule may be a suitable learning algorithm in the framework of temporal coding. The universal rule allows the rate-dependent potentiation to outweigh the STDP rule at sufficiently high firing rates in that the universal rule is approximated to a modified Hebb rule such as the BCM rule. To date, different models for the universal rule have been proposed to bridge the gap between spike- and firing rate-based learning rules.[67,69-72]

**C. Temporal difference learning and spike timing-dependent plasticity**

As indicated by Wörgötter and Porr[57], the TD learning rule in Eq. (10) shares some similar features with the synaptic modification rules in the previous section. In the TD learning, weight is modified only if non-zero $\delta$ and input $u$ are present. Notably, $\delta$ and $u$ are generated from an output and input node, respectively. These attributes are reminiscent of the above discussed synaptic modification rules in that the output and input node may correspond to a post and presynaptic neuron, respectively, and the required non-zero output and input signals for weight modification may indicate the temporal correlation between the nodes. In this regard, the TD learning may be linked to the STDP behavior.

The TD learning algorithm in SNN in Fig. 8 is simply on the ground of coincident input and delta spikes at the expense of an enormous number of auxiliary spiking neurons whose number proportionally increases with the interval between the input and reward spikes. The auxiliary neurons endow the input and output spikes with a temporal correlation. Alternatively, one can simply introduce a time-varying variable(s) in place of such auxiliary neurons, which provides a temporal correlation between the input and output spikes.

To this end, Rao and Sejnowski regarded the membrane potential of a postsynaptic neuron as the output (corresponding to $v$ in Eq. (10)) in the TD learning, which is a time-varying variable.[73] In this model, $\Delta t$ in the TD term $v(t+\Delta t) - v(t)$ was fixed to 10 ms so that weight increases if the postsynaptic membrane potential at $t+\Delta t$ is larger than that at $t$, i.e. $v(t+\Delta t) - v(t) > 0$. In contrast, if $v(t+\Delta t) - v(t) < 0$, the weight decreases in proportional to the magnitude of the TD term. Note that a reward is not given. The pre and postsynaptic neurons were forced to fire spikes with a temporal distance, either pre-post or post-pre order, and the consequent weight change was monitored. For the pre-post order, the presence of the postsynaptic spike causes a positive TD term due to a rise in the postsynaptic membrane potential given the postsynaptic spike, leading to a weight increase (potentiation). An example in this case is shown in Fig. 10a. Such potentiation is viable unless the distance between the spikes is so far that the preceding presynaptic spike barely perturbs the postsynaptic membrane. The opposite order however decreases the weight because $v(t+\Delta t) - v(t) < 0$ because the preceding postsynaptic spike elevates the membrane potential upon the spike and the potential decreases in due course. This implies depression. The spike distance also significantly influences the depression for the same reason as the potentiation. Fig. 10b is an example of such TD learning for post-pre order. The weight change was evaluated at different temporal distance between the pre and postsynaptic spikes as shown in Fig. 10c. Interestingly, the behavior is fairly similar to the STDP behavior, for instance, shown in Fig. 9. This may be evidence of a link between STDP and TD learning.

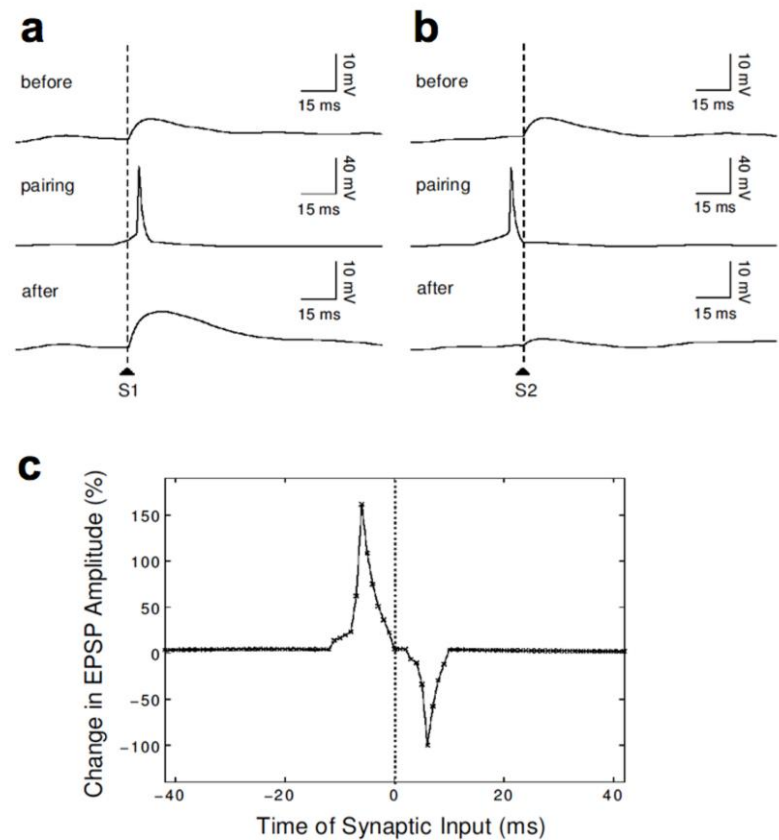

**Figure 10**. (**a**) Potentiation arising from TD learning algorithm upon pre-post paring. (**b**) Depression by the opposite spiking order. (**c**) The calculated weight change upon the difference in timing between the pre and postsynaptic spikes. Reproduced with permission.[73] Copyright 2001, MIT Press.

**V. Concluding remarks**

SNN as a time-dependent hypothesis is a unique function that has great potential to learn from time-varying input to figure out its temporal pattern. This learned temporal pattern is the substrate of future prediction and decision-making to maximize the utility. Yet, SNN and its learning algorithms are not sufficiently known to readily use the hypothesis in various temporal tasks, leaving a number of challenges with regard to practical application of SNN to temporal learning. However, this also points to opportunities that should be taken by pioneers. Fortunately, vigorous studies having been done for the last decades have enriched our understanding of brain functionalities and learning algorithms, which will hopefully be linked to SNN for temporal learning in the near future.

SNN may be viewed as a replacement for ANN within the framework of deep learning, thereby several SNN-suitable learning algorithms—functionally equivalent to the backpropagation algorithm—have been proposed. Such algorithms are excellently addressed in a recent review by Neftci.[74] In this framework, neuromorphic implementation of SNN offers a means of deep learning acceleration. Such an application specific integrated circuit likely achieves significant acceleration of training as well as inference in comparison with general-purpose hardware such as graphics processing units (GPUs) as for tensor processing units (TPUs)[75]. To this end, various attributes of neuromorphic chips as deep learning accelerators, e.g. performance (power consumption, speed, inference precision, and so forth) and cost, should thoroughly be considered compared with competing technologies that include general-purpose GPU, TPU, and FPGA.

Above all, breakthroughs in neuromorphic engineering are likely driven by its key applications that leverage its distinct capabilities from other technologies. Once again, capabilities of temporal learning and physical time domain operation make the neuromorphic system eligible for a unique learning platform. A good example is BrainScaleS,[76] a large-scale neuromorphic system, aiming to emulate large-scale SNN in accelerated time scale as an alternative to SNN simulation on general-purpose hardware. This SNN emulation platform can overcome the large gap between simulation runtime and physical time subject to simulation.

Neuromorphic system design for temporal learning is an important concern. Temporal learning varying in temporal distance between neighboring events may require rich dynamics that also varies in decay time constant. For instance, the TD learning with the aid of STDP in Section IV.C can only correlate events separated within the timing windows of STDP rule. This limit may be overcome by incorporating various learning dynamics with various time constants. Yet, enriching dynamics in neuromorphic system costs large power and circuit footprint; particularly, the use of digital arithmetic operators of high precision is a challenge. Additionally, the consequent need for large memory is an obstacle to large-scale integration of SNN.

Enriching dynamics using analog circuits is a conceivable solution to the aforementioned challenges in place of digital arithmetic operators. However, uncontrollable variability in devices, e.g. mismatch in circuit design, and the consequent variability in circuit behavior keep the scalability low. A feasible solution may be the use of mixed analog/digital circuits as for the neuron circuit in Braindrop.[6]

**Appendix: derivation of the delta learning rule for a linear network**

Suppose a linear neural network with $N$ nodes sharing another node through directed edges. The activity of $i$th node among $N$ nodes is denoted by $u_i$ ($i$ = 1, 2, …, $N$). A column vector $\boldsymbol{u}$ includes all $u_i$'s: $\boldsymbol{u}^\mathrm{T} = [u_1, u_2, …, u_N]$. The activity of the shared node is denoted by $v$. The given network topology yields $v$ as $v = \boldsymbol{w}^T \cdot \boldsymbol{u}$. $\boldsymbol{w}$ is a weight vector; $\boldsymbol{w} = [w_1, w_2, …, w_N]$. The task is to tweak $\boldsymbol{w}$ in order for $v$ to be close to a desired result $r$. Error-correction function $E$ is defined as $E = (r - v)^2/2$. Applying stochastic gradient descent (mini-batch size of one) yields the following component-wise update equation

$$w_i \rightarrow w_i - \eta \frac{\partial E}{\partial w_i} = w_i + \eta(r - v)u_i,$$

where $\eta$ is a learning rate. This equation is referred to as a delta learning rule given the presence of the delta term $\delta$ (=$r$ - $v$).

**Appendix: STDP in an activity domain**

Regarding the Poisson spike generation following a renewal process in nature, the probability that a presynaptic spike is elicited from presynaptic neuron $i$ during an infinitesimal time bin $dt_\mathrm{pre}$ is $a_i dt_\mathrm{pre}$, and for postsynaptic neuron $j$ $a_j dt_\mathrm{post}$. That is, the spiking probability has a uniform distribution over time unless $a_i$ and $a_j$ are time-variant. A weight change $\Delta w_{ji}$ given $t_\mathrm{pre}$ and $t_\mathrm{post}$ is expressed as Eq. (12).

Accordingly, the expected weight change $\langle \Delta w_{ji} \rangle$ during a training period $T$ (from $t_0$ to $t_0 + T$), where $a_i$ and $a_j$ are constant, is evaluated using the following equation,

$$\langle \Delta w_{ji} \rangle = \int_{t_0}^{t_0+T} \int_{-\infty}^{\infty} \Delta w_{ji}\, a_j dt_{post} a_i dt_{pre}. \tag{17}$$

Plugging Eq. (12) into this equation yields

$$\langle \Delta w_{ji} \rangle = a_i a_j \int_{t_0}^{t_0+T} \left[ A_+ \int_{t_{pre}}^{\infty} e^{-\frac{(t_{post}-t_{pre})}{\tau_+}} dt_{post} - A_- \int_{-\infty}^{t_{pre}} e^{\frac{(t_{post}-t_{pre})}{\tau_-}} dt_{post} \right] dt_{pre} =$$

$$(A_+ \tau_+ - A_- \tau_-) a_i a_j T. \tag{18}$$

To take into account the causality between pre and postsynaptic spikes, the spiking probability of postsynaptic neuron $j$ is assumed to be non-uniform as

$$p(t_{pre}, t_{post}) a_j dt_{post} + a_j dt_{post}, \tag{19}$$

where $p(t_{pre}, t_{post})$ is a non-uniform distribution function centered at $t_{pre} + \Delta t$ (>0). This distribution function is multiplied by $a_j$ to scale it with postsynaptic activity $a_j$ because it makes intuitive sense to observe more spikes around $t_{pre} + \Delta t$ with an increase in the postsynaptic activity. The second term in Eq. (19) indicates the background probability of postsynaptic spiking, which is constant ($a_j$) over the entire time. To this end, Eq. (18) can be re-written as

$$\langle \Delta w_{ji} \rangle = \int_{t_0}^{t_0+T} \int_{-\infty}^{\infty} \Delta w_{ji} [1 + p(t_{pre}, t_{post})] a_j dt_{post} a_i dt_{pre}$$

$$= \int_{t_0}^{t_0+T} \int_{-\infty}^{\infty} \Delta w_{ji}\, a_j dt_{post} a_i dt_{pre} + \int_{t_0}^{t_0+T} \int_{-\infty}^{\infty} \Delta w_{ji}\, p(t_{pre}, t_{post}) a_j dt_{post} a_i dt_{pre}, \tag{20}$$

The first term on the right-hand side is identical to Eq. (17), thereby giving the same result as Eq. (13). The second term can be simplified to $a_i a_j T l$ where $l$ is

$$l = A_+ \int_{t_{pre}}^{\infty} e^{-\frac{(t_{post}-t_{pre})}{\tau_+}} p(t_{pre}, t_{post}) dt_{post} - A_- \int_{-\infty}^{t_{pre}} e^{\frac{(t_{post}-t_{pre})}{\tau_-}} p(t_{pre}, t_{post}) dt_{post}.$$

Given that the distribution function $p$ is centered at $t_{\text{pre}} + \Delta t$, the first term on the right-hand side most likely outweighs the second term, yielding positive $l$. Therefore, Eq. (17) is eventually expressed as

$$\Delta w_{ji} = (A_+ \tau_+ - A_- \tau_- + l) a_i a_j T.$$

**References**

1 B. V. Benjamin, P. Gao, E. McQuinn, S. Choudhary, A. R. Chandrasekaran, J. M. Bussat, R. Alvarez-Icaza, J. V. Arthur, P. A. Merolla, and K. Boahen, Proceedings of the IEEE **102,** 699 (2014).

2 P. A. Merolla, J. V. Arthur, R. Alvarez-Icaza, A. S. Cassidy, J. Sawada, F. Akopyan, B. L. Jackson, N. Imam, C. Guo, Y. Nakamura, B. Brezzo, I. Vo, S. K. Esser, R. Appuswamy, B. Taba, A. Amir, M. D. Flickner, W. P. Risk, R. Manohar, and D. S. Modha, Science **345,** 668 (2014).

3 N. Q. S. Moradi, F. Stefanini, G. Indiveri, IEEE Transactions on Biomedical Circuits and Systems **12,** 106 (2018).

4 E. Painkras, L. A. Plana, J. Garside, S. Temple, F. Galluppi, C. Patterson, David R. Lester, and A. D. Brown, IEEE Journal of Solid-State Circuits **48,** 1943 (2013).

5 M. Davies, N. Srinivasa, T.-H. Lin, G. Chinya, Y. Cao, S. H. Choday, G. Dimou, P. Joshi, N. Imam, S. Jain, Y. Liao, C.-K. Lin, A. Lines, R. Liu, D. Mathaikutty, S. McCoy, A. Paul, J. Tse, G. Venkataramanan, Y.-H. Weng, A. Wild, Y. Yang, and H. Wang, IEEE Micro **38,** 82 (2018).

6 A. Neckar, T. C. Stewart, B. V. Benjamin, and K. Boahen, in *Optimizing an Analog Neuron Circuit Design for Nonlinear Function Approximation*, Florence, Italy, 2018 (IEEE).

7 C. Mead, *Analog VLSI and Neural Systems* (Adison-Wesley, Reading, MA, 1989).

8 C. Mead, Proceedings of the IEEE **78,** 1629 (1990).

9    G. Indiveri, B. Linares-Barranco, T. J. Hamilton, A. van Schaik, R. Etienne-Cummings, T. Delbruck, S.-C. Liu, P. Dudek, P. Häfliger, S. Renaud, J. Schemmel, G. Cauwenberghs, J. Arthur, K. Hynna, F. Folowosele, S. Saighi, T. Serrano-Gotarredona, J. Wijekoon, Y. Wang, and K. Boahen, Frontiers in Neuroscience **5,** 73 (2011).

10   F. S. E. Chicca, C. Bartolozzi, G. Indiveri, Proceedings of the IEEE **102,** 1367 (2014).

11   K. A. Boahen, Circuits and Systems II: Analog and Digital Signal Processing, IEEE Transactions on **47,** 416 (2000).

12   N. Qiao, H. Mostafa, F. Corradi, M. Osswald, F. Stefanini, D. Sumislawska, and G. Indiveri, Frontiers in Neuroscience **9** (2015).

13   A. Cassidy, A. G. Andreou, and J. Georgiou, in *Design of a one million neuron single FPGA neuromorphic system for real-time multimodal scene analysis*, 2011, p. 1.

14   J. Li, Y. Katori, and T. Kohno, Frontiers in Neuroscience **6,** 183 (2012).

15   V. Kornijcuk and D. S. Jeong, in *Pointer Based Routing Scheme for On-chip Learning in Neuromorphic Systems*, Rio de Janeiro, Brazil, 2018 (IEEE).

16   D. S. Jeong, I. Kim, M. Ziegler, and H. Kohlstedt, RSC Advances **3,** 3169 (2013).

17   D. S. Jeong and C. S. Hwang, Advanced Materials **0,** 1704729 (2018).

18   D. Kuzum, Y. Shimeng, and H. S. P. Wong, Nanotechnology **24,** 382001 (2013).

19   T. Stewart, B. Tripp, and C. Eliasmith, Frontiers in Neuroinformatics **3** (2009).

20   R. B. Stein, Biophysical Journal **5,** 173 (1965).

21   A. L. Hodgkin and A. F. Huxley, The Journal of Physiology **117,** 500 (1952).

22   S. Pepke, T. Kinzer-Ursem, S. Mihalas, and M. B. Kennedy, PLOS Computational Biology **6,** e1000675 (2010).

23   R. F. Oliveira, M. Kim, and K. T. Blackwell, PLOS Computational Biology **8,** e1002383 (2012).

24   T. Nakano, T. Doi, J. Yoshimoto, and K. Doya, PLOS Computational Biology **6,** e1000670 (2010).

25   J. Lisman, R. Yasuda, and S. Raghavachari, Nat Rev Neurosci **13,** 169 (2012).

26   R. S. Sutton, Machine Learning **3,** 9 (1988).

27   D. E. Rumelhart, G. E. Hinton, and R. J. Williams, Nature **323,** 533 (1986).

28   J. J. Hopfield, Proceedings of the National Academy of Sciences **79,** 2554 (1982).

29   J. B. Zachary C. Lipton, Charles Elkan, arXiv:1506.00019 [cs.LG] (2015).

30   S. Hochreiter and J. Schmidhuber, Neural Computation **9,** 1735 (1997).

31   C. Eliasmith and C. H. Anderson, *Neural Engineering: Computation, Representation, and Dynamics in Neurobiological Systems* (MIT Press, Cambridge, MA, 2003).

32   J. C. Eccles, P. Fatt, and K. Koketsu, The Journal of Physiology **126,** 524 (1954).

33   N. Uchida, Nature Neuroscience **17,** 1432 (2014).

34   W. Gerstner and W. M. Kistler, *Spiking Neuron Models: Single Neurons, Populations, Plasticity* (Cambridge University Press, 2002).

35   P. Dayan and L. F. Abbott, *Theoretical Neuroscience* (The MIT Press, London, 2001).

36   E. M. Izhikevich, IEEE Transactions on Neural Networks **14,** 1569 (2003).

37   R. FitzHugh, Biophysical Journal **1,** 445 (1961).

38   J. Nagumo, S. Arimoto, and S. Yoshizawa, Proceedings of the IRE **50,** 2061 (1962).

39   P. J. Sjöström and M. Häusser, Neuron **51,** 227 (2006).

40   P. N. Steinmetz, A. Manwani, C. Koch, M. London, and I. Segev, Journal of Computational Neuroscience **9,** 133 (2000).

41   J. A. White, J. T. Rubinstein, and A. R. Kay, Trends in Neurosciences **23,** 131 (2000).

42   A. A. Faisal, L. P. J. Selen, and D. M. Wolpert, Nature Reviews Neuroscience **9,** 292 (2008).

43   B. Katz and R. Miledi, Nature **226,** 962 (1970).

44   M. N. Shadlen and W. T. Newsome, Current Opinion in Neurobiology **4,** 569 (1994).

45   C. van Vreeswijk and H. Sompolinsky, Science **274,** 1724 (1996).

46   D. Debanne, Nature Reviews Neuroscience **5,** 304 (2004).

47   J. G. Jefferys, Physiological Reviews **75,** 689 (1995).

48   G. Maimon and J. A. Assad, Neuron **62,** 426 (2009).

49   D. Heeger, "Poisson model of spike generation," (2000).

50   T. M. Cover and J. A. Thomas, *Elements of Information Theory*, 2nd ed. (Wiley-Interscience, 2006).

51  E. T. Rolls, A. Treves, and M. J. Tovee, Experimental Brain Research **114,** 149 (1997).

52  E. T. Rolls, L. Franco, N. C. Aggelopoulos, and J. M. Jerez, Vision Research **46,** 4193 (2006).

53  F. Theunissen and J. P. Miller, Journal of Computational Neuroscience **2,** 149 (1995).

54  E. D. Adrian and Y. Zotterman, The Journal of Physiology **61,** 151 (1926).

55  W. Bialek, F. Rieke, R. de Ruyter van Steveninck, and D. Warland, Science **252,** 1854 (1991).

56  Simon J. Thorpe, in *Parallel processing in neural systems and computers*, edited by R. Eckmiller, G. Hartmann, and G. Hauske (Elsevier, New York, NY, USA, 1990), p. 91.

57  F. Wörgötter and B. Porr, Neural Computation **17,** 245 (2005).

58  P. Montague, P. Dayan, and T. Sejnowski, The Journal of Neuroscience **16,** 1936 (1996).

59  D. O. Hebb, *The organization of behavior* (Wiley & Sons, New York, 1949).

60  E. Bienenstock, L. Cooper, and P. Munro, The Journal of Neuroscience **2,** 32 (1982).

61  L. N. Cooper and M. F. Bear, Nat Rev Neurosci **13,** 798 (2012).

62  L. N. Cooper, F. Liberman, and E. Oja, Biological Cybernetics **33,** 9 (1979).

63  G.-q. Bi and M.-m. Poo, The Journal of Neuroscience **18,** 10464 (1998).

64  S. Song, K. D. Miller, and L. F. Abbott, Nat. Neurosci. **3,** 919 (2000).

65  P. J. Sjöström, G. G. Turrigiano, and S. B. Nelson, Neuron **32,** 1149 (2001).

66  S. M. Dudek and M. F. Bear, Proceedings of the National Academy of Sciences **89,** 4363 (1992).

67  J. Gjorgjieva, C. Clopath, J. Audet, and J.-P. Pfister, Proceedings of the National Academy of Sciences **108,** 19383 (2011).

68  V. Kornijcuk, H. Lim, I. Kim, J.-K. Park, W.-S. Lee, J.-H. Choi, B. J. Choi, and D. S. Jeong, Scientific Reports **7,** 17579 (2017).

69  E. M. Izhikevich and N. S. Desai, Neural Computation **15,** 1511 (2003).

70  H. Z. Shouval, M. F. Bear, and L. N. Cooper, Proceedings of the National Academy of Sciences **99,** 10831 (2002).

71  J.-P. Pfister and W. Gerstner, The Journal of Neuroscience **26,** 9673 (2006).

72   C. Clopath, L. Büsing, E. Vasilaki, and W. Gerstner, Nature Neuroscience **13,** 344 (2010).

73   R. P. N. Rao and T. J. Sejnowski, Neural Computation **13,** 2221 (2001).

74   E. O. Neftci, iScience **5,** 52 (2018).

75   N. P. Jouppi, C. Young, N. Patil, D. Patterson, G. Agrawal, R. Bajwa, S. Bates, S. Bhatia, N. Boden, A. Borchers, R. Boyle, P.-l. Cantin, C. Chao, C. Clark, J. Coriell, M. Daley, M. Dau, J. Dean, B. Gelb, T. V. Ghaemmaghami, R. Gottipati, W. Gulland, R. Hagmann, C. R. Ho, D. Hogberg, J. Hu, R. Hundt, D. Hurt, J. Ibarz, A. Jaffey, A. Jaworski, A. Kaplan, H. Khaitan, D. Killebrew, A. Koch, N. Kumar, S. Lacy, J. Laudon, J. Law, D. Le, C. Leary, Z. Liu, K. Lucke, A. Lundin, G. MacKean, A. Maggiore, M. Mahony, K. Miller, R. Nagarajan, R. Narayanaswami, R. Ni, K. Nix, T. Norrie, M. Omernick, N. Penukonda, A. Phelps, J. Ross, M. Ross, A. Salek, E. Samadiani, C. Severn, G. Sizikov, M. Snelham, J. Souter, D. Steinberg, A. Swing, M. Tan, G. Thorson, B. Tian, H. Toma, E. Tuttle, V. Vasudevan, R. Walter, W. Wang, E. Wilcox, and D. H. Yoon, arXiv:1704.04760v1 (2017).

76   J. Schemmel, D. Briiderle, A. Griibl, M. Hock, K. Meier, and S. Millner, in *A wafer-scale neuromorphic hardware system for large-scale neural modeling*, 2010, p. 1947.

**Acknowledgements**

This work was supported by the research grant (2018K2A9A2A08000151) of the National Research Foundation of Korea (NRF).